\documentclass[journal=jacsat,manuscript=letter]{achemso}

\usepackage[version=3]{mhchem} 
\usepackage[T1]{fontenc}       
\usepackage{xspace}
\usepackage{multirow}
\usepackage[usenames,dvipsnames,svgnames,table]{xcolor}
\usepackage{soul} 

\newcommand*\abinit{\textit{ab initio}\xspace}
\let\oldAA\AA
\renewcommand*{\AA}{\oldAA\xspace}

\author{R\'{e}mi Khatib}
\affiliation[JGU]{Department of Physics, Johannes Gutenberg University, Staudingerweg 7, 55128 Mainz, Germany}
\author{Ashwinee Kumar}
\affiliation[TCD]{School of Physics, Advanced Materials and Bioengineering Research Centre (AMBER) and Centre for Research on Adaptive Nanostructures and Nanodevices (CRANN) Institute, Trinity College, Dublin 2, Ireland}
\author{Stefano Sanvito}
\affiliation[TCD]{School of Physics, Advanced Materials and Bioengineering Research Centre (AMBER) and Centre for Research on Adaptive Nanostructures and Nanodevices (CRANN) Institute, Trinity College, Dublin 2, Ireland}
\author{Marialore Sulpizi}
\affiliation[JGU]{Department of Physics, Johannes Gutenberg University, Staudingerweg 7, 55128 Mainz, Germany}
\author{Clotilde S. Cucinotta}
\affiliation[ICL]{Department of Chemistry, Imperial College London, W12 0BZ, UK and TYC}
\email{c.cucinotta@imperial.ac.uk}
\phone{+44 20 759 41169 }

\title{The nanoscale structure of the Pt-water double layer under bias revealed}

\abbreviations{DFT}

\begin{document}

\begin{abstract}
The nanoscopic mass and charge distribution within the double layer at electrified interfaces plays a key r\^ole in electrochemical phenomena of huge technological relevance for energy production and conversion. However, in spite of its importance, the nanoscopic structure of the double layer and its response to an applied potential is still almost entirely unknown, even for Pt-water, the most fundamental electrochemical interface. Using a general \abinit methodology which advances previous models towards a dynamic and more realistic description of an electrode/electrolyte interface, we simulate for the first time the nanoscopic structure of the Pt-water double layer and its response to an applied potential, in realistic solution conditions. We reveal that the nanoscopic metal/surface structure and charging are not captured by traditional capacitor models, as the electrode polarization is associated with a charge oscillation within the double layer and a densification of the water layer in contact with the electrode, both of which strongly depend on the applied potential. Furthermore, we demonstrate that the interface dipole is not determined by the reorientation of the first water layer in contact with the electrode, but by its charging state in combination with its number density, while water reorientation becomes relevant only in the second water layer. Our findings will be essential to develop highly realistic models for the catalytic processes at the Pt-water interface.
\end{abstract}


The last decades have witnessed an extensive research effort aimed at expanding energy production, transformation and storage, through the design and optimisation of a variety of electrochemical (EC) devices. These include fuel cells, solar cells, super capacitors and batteries. These devices could potentially enable a cleaner, cheaper and safer energy technology, but their performance is still not adequate for commercial applications and their development is slow. The fundamental reason behind this is that the optimization of these devices still proceeds by trial and error, due to a lack of atomistic understanding of the basic processes across the electrified interfaces (EI)s which underlie the functioning of such devices. The formation of a double layer (DL) is a fundamental process occurring at EIs. Its equilibrium nano-structure and its response to an applied potential strongly affect the activation barriers for the mass and charge transfer at EIs which converts chemical into electrical energy and \textit{vice versa}. 

Modern experimental techniques,  such as scanning tunnel microscopy (STM), atomic force microscopy 
(AFM) and near field spectroscopy (NFS), can achieve 
atomic resolution. However, in most experimental situations it is only possible to observe the macroscopic manifestation of EC conversion processes. Therefore, in spite of the vast amount of empirical data on EC systems accumulated since Faraday's ground-breaking work~\cite{Faraday1834}, there is still a wide gap between experimental observation of EC processes and their microscopic interpretation.

On the theoretical side, computer simulations can achieve atomistic and electronic resolution and are therefore essential to developing better devices. However, despite a long history of empirical~\cite{Raghavan1991,Heinz2008,Corni2009} and \abinit~\cite{Cicero2008,Meng2002,ZeroChargeJiabo,Finnis2018}
theoretical surface science~\cite{Schnur2009,Roman2013, Tanblyn2014,Todorova2014}
investigations~\cite{Rossmeisl2005, ROSSMEISL200868}, 
resulting in deep insight into the structure and reactivity of interfaces~\cite{Schmickler2010},
a comprehensive mechanistic description of atomic and electronic scale processes at electrode
surfaces under applied potential is still lacking, even for the most fundamental interfaces, such as the Pt-water interface.

In order to describe the EC processes at EIs it is necessary to model the effect of the application of a potential to the cell, since the electrode potential, U, is an essential parameter in electrochemistry. 
A milestone in the modelling of the effect of the EC potential is represented by the work of 
N\o rskov, who in 2004, using the fact that for a standard (pH=0, p=1 bar, T=298 K) hydrogen electrode (SHE), U=0 is defined by the equilibrium condition
\begin{align}
    \ce{H_2(g)  <=> H^{+}(aq) + e^{-}},\:
\end{align}
demonstrated~\cite{Norskov2004} that an implicit incorporation of the applied potential to a cell can be modeled. This incorporation is realised through the concept of the computational hydrogen electrode, where a computational reference to the SHE is established by imposing condition (1), without explicitly calculating any solvation energies.
N\o rskov's method has been successful in predicting trends in various EC reactions~\cite{Norskov2004,Norskov2005}. However, in his method the electrode potential only affects the Fermi level of the electrons, and the EC environment is only considered as the adsorbates' reservoir. The explicit representation of the atomistic structure of the EI, and its effect on adsorption energetics, as well as the effect of the applied potential on the interfacial geometry, charge redistribution and electronic structure, are neglected. These are all crucial components of the EC phenomenon, whose modelling is essential for a realistic description of electrode activity~\cite{Staszak2015}. A more realistic method is therefore needed.

More recent simulation methods explicitly addressing the description of the EI, typically use supercell geometries and periodic boundary conditions (PBC) in three dimensions to speed up the calculations and minimise the effect of finite sample sizes. Since PBC make it difficult to apply an electrostatic potential to the cell, this effect is modelled by imposing charges on the electrodes; cell neutrality is achieved by adding a distribution of fixed counter-charges to the cell, which can take the form of a homogeneous background filling the simulation cell~\cite{Filhol2006}, of charged planes~\cite{Lozovoi2001}, a doped semiconductor electrode~\cite{Finnis2018} or hydronium ions placed at a fixed distance from the electrode surface~\cite{Norskov2004}. 
  Most of these methods reproduce the localised electric 
field and the potential energy drop within a nanoscopic distance from the metal surface~\cite{Filhol2006}.
However, none of them has provided a comprehensive atomistic description of interface structure
and charge polarisation within the DL under applied potential. This is not only due to the small size of the samples used; it is also due to the lack of an extended dynamic description of the DL nano-structure based on first principles that can describe the liquid nature of the electrolyte and the emergence of huge interfacial electric fields.
A more realistic modelling of the EC environment, in combination with well controlled experiments, is crucial to develop more reliable models for catalytic processes at EIs~\cite{Subbaraman2011,Staszak2015}.

An alternative computational SHE method
 based on \abinit molecular dynamics (AIMD), but relying on an approach that might be seen as intermediate between simulation and theory,
    was developed by Sprik and co-workers~\cite{Cheng2009,Costanzo2011,Cheng2014,ZeroChargeJiabo}. This method makes it possible to refer the computed potential to the SHE by using the solvation free energy of the proton in water. 
   The method is sophisticated and powerful, but one downside is its high computational cost. 
  Furthermore, the method does not provide a description of the over-potential associated to the transfer of species to, from, or across the interface under bias, and does not include an explicit description of the electrolytic solution or address the problem of controlling the charging of the electrode. 

\begin{figure}
 	\includegraphics[height=7.5cm]{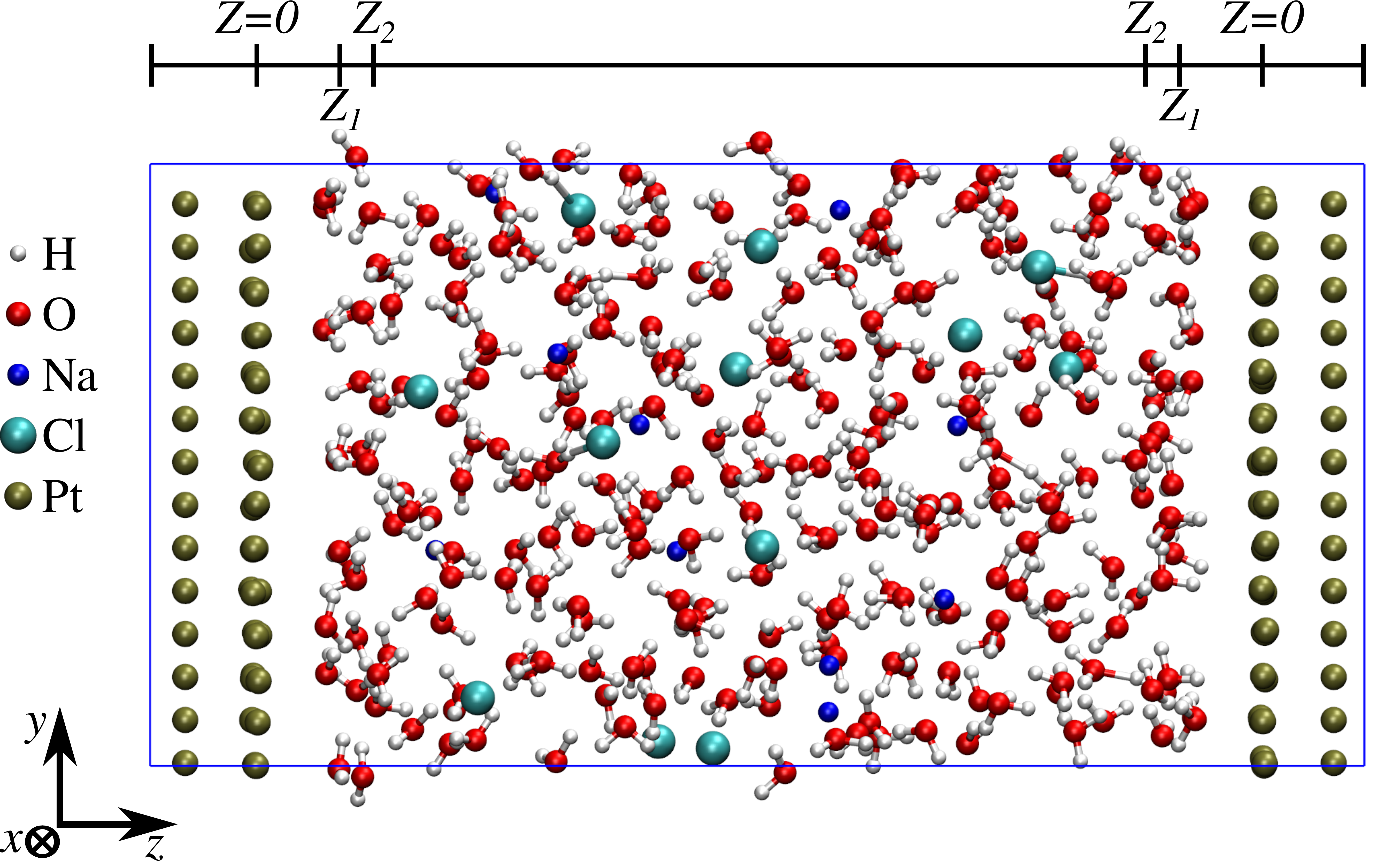}
 	\caption{A snapshot taken from the AIMD trajectory of the Pt-water solution system characterised by an electrode charge of -0.03 |e|, and 10 Na and 12 Cl ions in solution (10Na:12Cl). The horizontal bar on the top of the figure marks the external boundary of the first and second water layer ($Z_1$ and $Z_2$, respectively, as per Tab. 1 in SI). $Z=0$ labels the average position of the surface Pt layers in contact with the aqueous electrolyte. Red, white, olive green, cyan and blue spheres represent O, H, Pt, Cl and Na atoms, respectively.}
 	\label{fig:snapshot}
 \end{figure}

For these reasons we abandon part of the theoretical sophistications of the previous methods and adopt a simpler, intuitive description of the electrode/electrolyte interface. We here use AIMD simulations to carry out a virtual experiment and directly reproduce the nanostructure of a Pt-water EI under applied potential, in realistic solution conditions (Fig.~\ref{fig:snapshot}). We evaluate the capacitance and the potential of zero charge of this interface as well as the electrode potentials (wrt. the SHE) associated to different charging states of the Pt electrode. We reproduce the Sum Frequency Generation (SFG) spectrum for this interface.

To achieve this, we develop a highly realistic, general \abinit methodology, the \textit{ion unbalance} methodology, which enables the direct simulation of a metal/water EI under bias and  accounts explicitly for charge polarisation effects at both sides of the EI, the full dynamics of solvent rearrangement and the electronic structures details. 

\section{Results and Discussion}
\subsection{The \textit{ion unbalance} methodology}
  The \textit{ion unbalance} methodology enables the direct observation of the effect of an applied potential on the nanostructure of an EI under bias, by simulating and comparing EIs with the electrode in different charging states. 
Different charging states of the electrode are obtained by introducing different unbalanced populations of neutral atoms of different types (Na and Cl) in the electrolyte solution. This introduction then leads to the formation of cations and anions in solution and a charged electrode. 

In \abinit simulations, the density functional theory (DFT) approach determines \textit{self-consistently} the charge of each species by filling the Kohn-Sham electron states according to their relative energy with respect to the system's Fermi energy, which is in turn determined by the metal.
If the metal electrode is large enough to act as a suitable charge reservoir, the charge of each ion will
be determined by the transfer of electrons to/from the electrode, such that overall charge neutrality will be maintained.
The introduction of different imbalances in the number of cations and anions in the solution will thus result in the transfer of different amounts of charge between the electrolyte and the Pt electrode.
For instance, adding an imbalance between atomic species to the solution leading to the formation of more cations than anions will result in a more negatively charged electrode and an excess of cations in solution. These ionic species in solution will, in turn, contribute to the DL screening of the electrode.

\begin{figure}
  \includegraphics[scale=0.35]{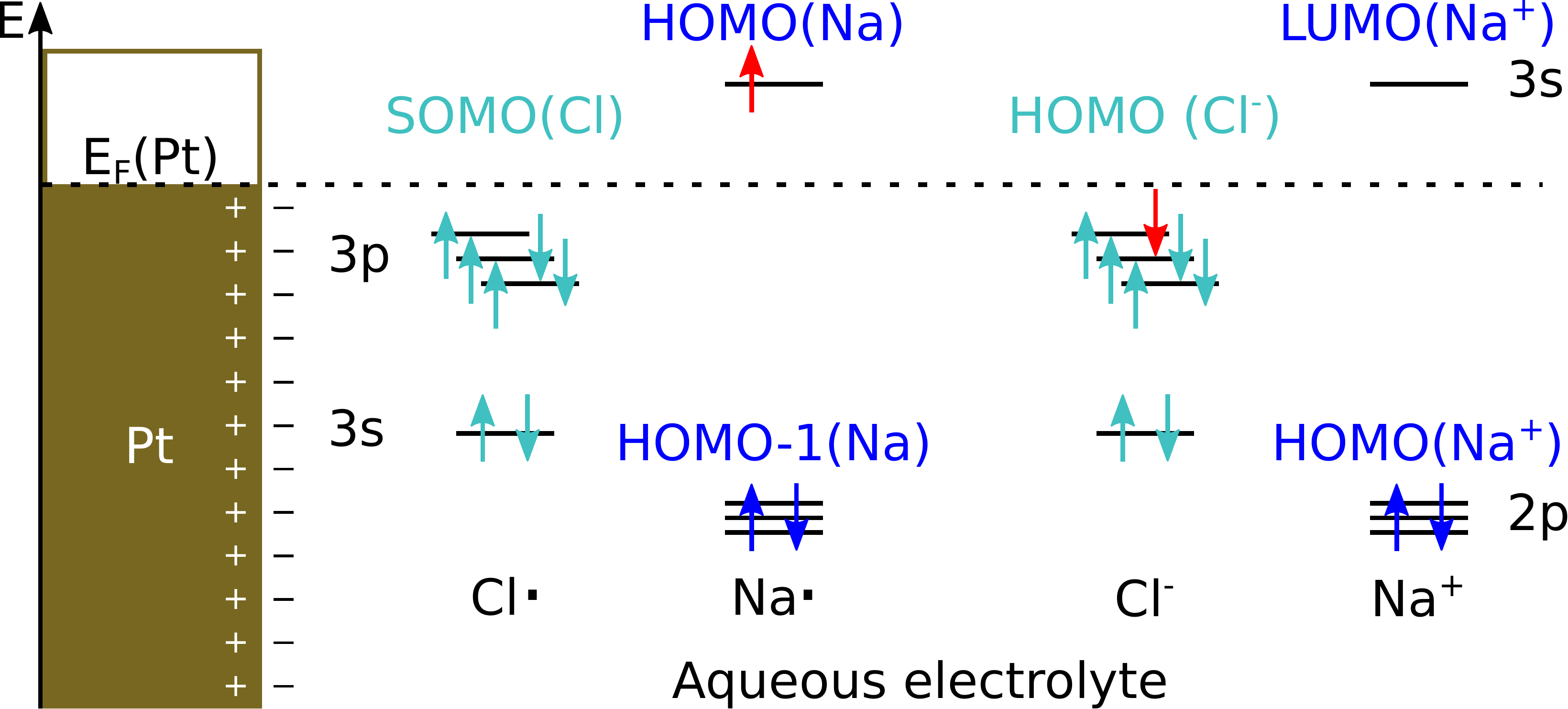}
  \caption{ Single electron energy level alignment and schematic picture of a Pt-electrolyte half-cell. $E_{F}(\mathrm{Pt})$ separates filled and empty electronic states. On the right, the equilibrium condition for the occupation of the energy levels of Na+ and Cl- ions in solution. On the left, the initial condition for the occupation of the energy levels centred around Na and Cl atomic species immediately after they are immersed in solution and before the charge transfer from the electrode to the electrolyte occurs. The highest occupied state for Na in solution, here HOMO(Na) (by analogy with the nomenclature for highest occupied molecular orbitals) is above $E_F$, thus it will be empty and Na species is expected to be fully ionised. Correspondingly, the lowest unoccupied state for Cl in solution, here SOMO(Cl) (the degenerate semi occupied molecular orbital) is below $E_F$, thus it will be filled and Cl species is expected to be negatively charged.Filled and empty water molecular states lie well below and above $E_{F}$, respectively. Note that a frozen picture of the single electron energy levels in our system is adopted here.}
 \label{fig:Levels}
\end{figure}

The validity of the proposed \textit{ion unbalance} methodology to charge the electrode, relies on the correct DFT description of the single electron energy 
level alignment, which may change for different approximations of the exchange and correlation functional. 
Correct energy level alignment should reflect the ordering shown in Fig.~\ref{fig:Levels}. We find that in our Pt-water interface this is achieved already by the generalised gradient approximation at the level of the Perdew-Burke-Ernzerhof~\cite{PBE} (PBE) potential~\cite{ZeroChargeJiabo}, despite the fact that other properties, such as the water band gap and the band gap of several semiconductors may be severely underestimated~\cite{galli2004}.

In the \textit{ion unbalance} methodology, the electrostatic potential drop for each electrode charging state (i.e. each unbalance in the population of ions) is obtained \textit{a posteriori}, from the corresponding equilibrium charge density distribution. Different potential drops can be associated to electrodes at different potentials, measurable e.g. with respect to the SHE. 
The bulk electrolyte energy level, which is located in the middle of each electrolyte region, represents a common reference for the systems in different electrode states, and will be used to align the respective electronic structures, eliminating the need for artificially introducing any vacuum slabs in the middle of the cell\cite{Filhol2006}. Note that to exploit the \textit{ion unbalance} methodology proposed, we need the aqueous electrolyte to approach bulk conditions in its central region, therefore system's size and the simulation times are highly relevant in our methodology.

Our methodology uses PBC approach. An advantage of this approach is that it will produce a single type of interface (the \textit{half-cell}), even in the case of the charged electrodes, since the screening between the two electrode surfaces in the cell will decouple them and the excess of ions will be equally partitioned between the two electrode surfaces. This makes it possible both to disentangle  the r\^ole of anode and cathode, and to improve the statistical sampling  from the AIMD simulations, by averaging over the two interfaces present in the simulation cell. 

Overall, with our methodology we provide a realistic framework which will enable the \abinit estimation of the internal energy, the entropy and the free energy along the reaction paths for chemical transformation at EIs, simulated in a realistic charged environment. This estimate can also be used to develop and tune semi-empirical models for the simulation of the steady state of large systems over time scales longer than those currently achievable with fully \abinit simulations.

\subsection{The model system}
In particular, in our model system (see Fig.~\ref{fig:snapshot}) a Pt slab in contact with a NaCl aqueous solution is realistically represented by about 1500 atoms and 5250 electrons. The entire model system is neutral and there is no charge background. We simulate the \abinit dynamics of three interfaces of nearly equivalent solvent composition ($\approx$ 2 M), differing only slightly in the relative number of Na and Cl ions included in the solution. We name these interfaces after their solvent compositions, thus, 10Na:10Cl, 12Na:10Cl and 10Na:12Cl.

The system size is adequate for our purposes for a number of reasons. Firstly, at the electrolyte concentration considered ({$\approx$}2~M of NaCl in water) the separation between the two electrode surfaces ($d$=28 \oldAA) corresponds to {$\approx$}8 Debye screening lengths. This reduced Debye length ensures that the electrolyte bulk condition is reached in the centre of the cell, i.e. at a relatively small distance from the surface of the electrode. Secondly, the large cell cross section (286~\oldAA$^2$) reduces the effects of the 2D periodicity along the electrode surface, justifying the minimal $k$-point sampling ($\Gamma$-point only) of the system Brillouin zone. Thirdly, the Pt slab alone contains nearly 3000 electrons. Thus it represents a sizeable charge reservoir able to correctly position the system Fermi energy and to stabilise its value even in the case of transfer of charge from the electrolyte to the electrodes.
Finally, the overall size of the model system allows us to accumulate sufficient statistics even during the relatively short timescales accessible to \abinit simulations. We have simulated each of the charged systems for the unprecedented time of {$\approx$}50~ps after equilibration. Overall, more than 200 ps of Born Oppenheimer AIMD have been simulated on our systems.  Note that the AIMD simulation of this huge, yet relatively simple model system constitutes a substantial computational challenge, well beyond current standard, which was only managed due to the efficient implementation of DFT based AIMD in CP2K code\cite{quickstep}.

In the following, we demonstrate how our methodology and model system make possible to overcome the traditional picture for the charge and mass distribution at the electrode-electrolyte interface and obtain a direct observation of the strong dependence of DL charge, water structure, density and orientation within the DL on an applied potential.

\subsection{The new picture for the Pt-water DL under applied potential}
In the conventional Gouy-Chapman-Stern picture of the DL, the electrode charge is strictly localised on the metal electrode, and is counterbalanced by the charge of the counter ions adsorbed on the metal surface and those present in the diffuse layer, the region with an ion concentration gradient.
The first feature of the Pt-water DL revealed by our methodology is that the surface charging at the metal-water solution interface cannot be simply described in terms of the Gouy-Chapman-Stern capacitor model.
 A more complex picture emerged from our calculations, where
(1) electrode charge spills over the water molecules in contact with the metal, therefore the charge on the electrolyte moiety of the DL is not determined uniquely by the counter ions; 
(2) an oscillating charge distribution is present at the metal-water interface;
(3) electrode polarization is associated with a change in the number of molecules in the first water layer in contact with the electrode, and a reorientation of the molecules of the second water layer.
(4) the interfacial charge and molecular distribution strongly depends on the applied potential. 

Within this picture the electrode charge is not exclusively localised on the electrode moiety of the interface, but includes the electronic charge on the electrode and in the water molecules in its proximity, whose number changes as a function of the applied potential. 
This picture is consistent with and provides the computational baseline for the general definition of the \textit{electrode total charge} concept, which was developed by Frumkin~\cite{Frumkin1980} based on thermodynamic considerations, and without relying on any traditional model for the DL. In this thermodynamic definition, rather then in terms of localisation on the electrode, charge is defined operationally, as the amount of electricity which needs to be supplied to the electrode when its surface is increased by 1 $cm^2$, and the concentration of all the solution components remains constant~\cite{Frumkin1980}. 

\subsubsection{The charge distribution}
A quantitative foundation to the new model for the DL nano-structure emerged from our calculations is first of all provided by the analysis of the excess (nominal - calculated) Bader charges distribution across the interface.

\begin{table}
	\caption{Bader excess (nominal-calculated) valence charges in units of |$e$|. In column `System', the name of the interface under consideration, defined in terms of the number of \ce{Na+} and \ce{Cl-} ions in solution and the type of metal electrode. 
		In columns 2-6 the total, subsurface and external surface charge, and the charge on first and second layer of water in contact with the electrode. These values are averaged along the trajectory and between the two interfaces present in our cell. Columns ``Cl + H`` and ``Na`` report the charge localised around Cl and Na ions (see SI). More specifically, ``Cl + H`` reports the sum of the electronic charge around Cl plus some charge artificially associated to the H atoms surrounding Cl (due to the crude Bader definition of atomic volume boundaries, part of the charge actually around Cl is artificially associated by the Bader scheme to the H atoms pointing towards the Cl anion). The last two columns ``Solvation Shells`` and ``DL``, report the total charge in the first solvation shells of both Na and Cl ions, and the total charge on the double layer, which includes the charge on electrode, first and second water layer.  A full picture of Bader charge distribution is reported in the SI.}
	\label{tbl:all}
	\resizebox{\textwidth}{!}{
		\begin{tabular}{|c|c|c|c|c|c|c|c|c|c|}
			\hline
			\multirow{2}{*}{System} &
			\multicolumn{3}{c|}{Metallic electrode} &
			\multicolumn{2}{c|}{Interfacial water} &
			\multicolumn{2}{c|}{Ions}  &
			\multicolumn{1}{c|}{Solvation Shells} &
			\multicolumn{1}{c|}{DL} \\
			\cline {2-8}
			&Total           &Subsurface       &Surface       &Layer 1          &Layer 2        &Cl+H           &Na     & &      \\
			\hline
			10Na:12Cl-Pt   &-0.03$\pm$0.08   &1.67$\pm$0.05 &-1.70$\pm$0.06   &0.77$\pm$0.12  &-0.08$\pm$0.06  &-0.83$\pm$0.01  &0.9109$\pm$0.0004   &-1.96    & 0.65 \\
			10Na:10Cl-Pt   &-0.61$\pm$0.12   &1.68$\pm$0.05 &-2.29$\pm$0.08   &0.49$\pm$0.18  &-0.16$\pm$0.74  &-0.80$\pm$0.01  &0.9049$\pm$0.0007   &0.14     &-0.28 \\
			12Na:10Cl-Pt   &-1.10$\pm$0.07   &1.70$\pm$0.04 &-2.80$\pm$0.05   &0.29$\pm$0.10  &-0.23$\pm$0.06  &-0.81$\pm$0.01  &0.9090$\pm$0.0002   &1.50     &-1.04 \\
			0Na:0Cl-Pt     &-0.56$\pm$0.08   &1.65$\pm$0.05 &-2.21$\pm$0.08   &0.62$\pm$0.08  &-0.09$\pm$0.07  &      0         & 0                  &  0      &  0   \\
			\hline
		\end{tabular}
	}
\end{table}

Firstly, Bader analysis confirms the viability of the \textit{ion unbalance} methodology, showing that the electrode state can be effectively controlled by the relative unbalance in the population of anions and cations in solution. Indeed, in our calculations a larger excess of positive (negative) ions in solution always corresponds to a more negatively (positively) charged electrode (see Tab.~\ref{tbl:all} and SI).
 
Secondly, our Bader analysis shows that the charge on the electrode moiety of the DL is mostly localised on the external electrode's surface layers and is zero in the centre of every metal slab. We find that the charge on the surface layer of the Pt slab strongly depends on the number of ions in solution, whilst the charge in the subsurface layer of the slab is the same for all interfaces under consideration (see {``}Surface{''}  and {``}Subsurface{''} in Tab.~\ref{tbl:all}). Note that the Pt atoms of the subsurface layers are kept fixed during the siimulation. 

\begin{figure}
	\includegraphics[scale=0.8]{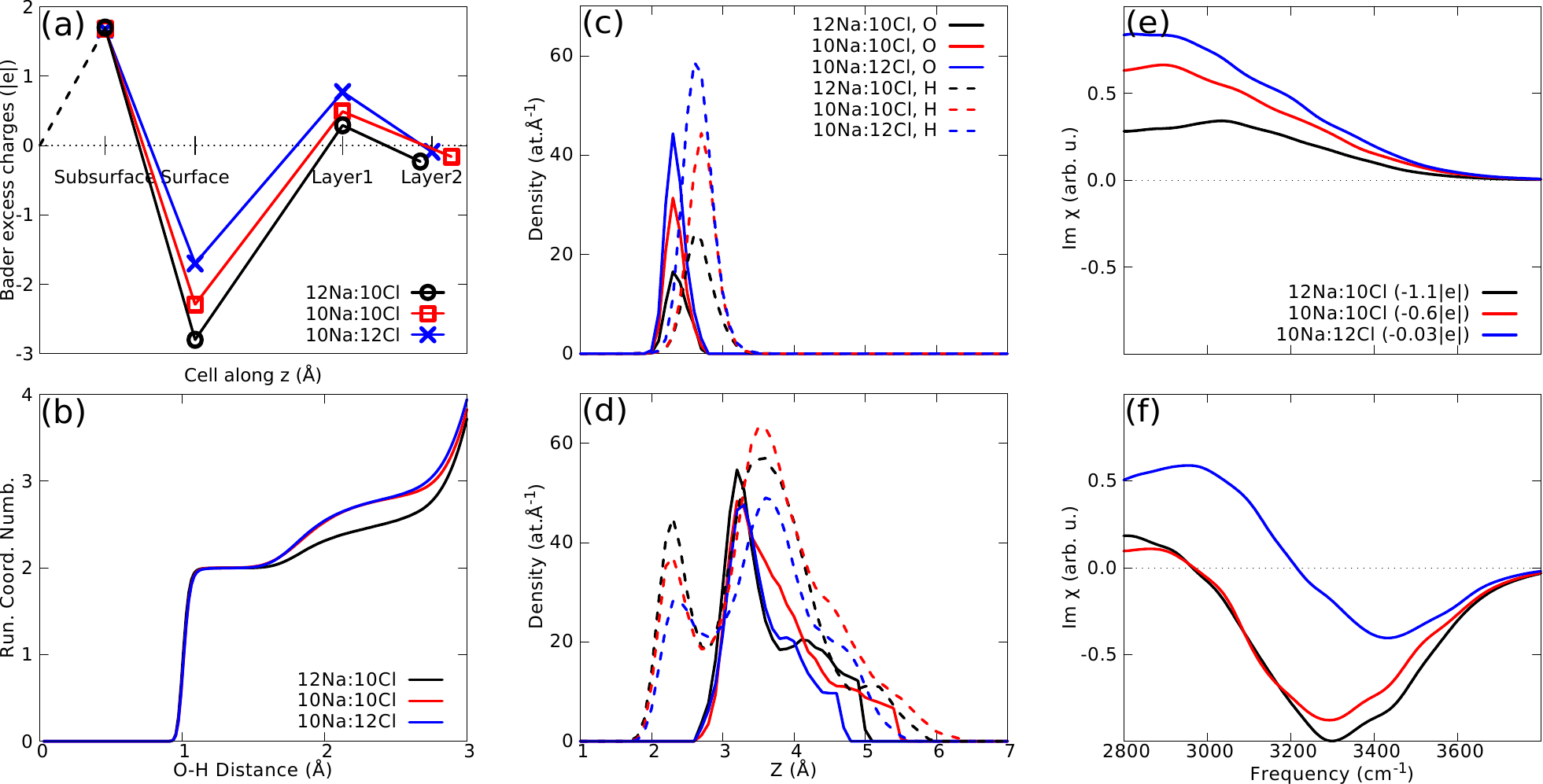}
	\caption{(a) Overall excess (nominal-calculated) Bader valence charge distribution 
		along the double layer. The black dashed line (joining the Pt subsurface layers and the region in their middle) signals that the in the middle of the metal slab the actual charge is 0 (see Fig.~1 in SI); (b) integrated O-H radial distribution profle, where O belongs to the molecules of the frst water layer in contact with the electrode; (c-d) average atomic density profiles for the O and H atoms belonging to the first (c) and second (d) water layers in contact with the electrode (in units of atoms per \AA); The plain and dashed lines stand for the O and H atoms' contributions, respectively; (e-f) layer resolved aurface sensitive vibrational density of states (VDOS) obtained only from the molecules of first (e) and second (f) water layer. The charge on the metal moiety of each DL is reported in brackets, in units of |$e$|. 12Na:10Cl, 10Na:10Cl, and 10Na:12Cl systems are represented in black, red and blue, respectively. For each system, all the represented quantities are averaged over 50 ps long trajectories at 330K, obtained performing DFT based Born Oppenheimer AIMD. First and second water layers are defined as described in Tab. 1 in SI.}
	\label{fig:combined}
\end{figure}

Thirdly, moving out from the negatively charged metal surface, we encounter a net positive electronic excess charge in the 
first water layer in contact with the electrode and a negative electronic excess charge in the second water layer, before reaching 
neutrality again in the bulk electrolyte, in the central part of the cell (see Tab.~\ref{tbl:all} and  
Fig.~\ref{fig:combined}(a)). 

The observed oscillating trend in the charge distribution at the DL is qualitatively the same for all interfaces studied; however, the amount of charge on each water layer is strongly dependent on the electrode state.
Note that the electronic charge spillover to the first water layer in contact with the electrode occurs such that each molecule always carries the same amount of positive charge ($\approx0.1 $~|$e$|), irrespective of the electrode state. Our calculations show that the number of equally charged molecules composing the first water layer progressively increases when the electrode becomes more positive (see Tab. 2 in SI). Therefore, the overall charge on the water layer in contact with the differently charged electrodes is determined by the different number of equally charged molecules composing this layer (see Tab.~\ref{tbl:all} and next paragraph). This means that the electrode charge is screened by changing the water density at the interface.

Overall, we find that the negative charge on the Pt electrode and within the DL is larger than that expected from traditional models~\cite{Rossmeisl2007} (see SI). In particular, the Bader charge on the electrode side of the DL amounts to -1.1$e$, -0.6$e$ and -0.03$e$ and the overall charge on the DL amounts to 0.65$e$, -0.28$e$ and 1.5$e$, for the 12Na:10Cl, 10Na:10Cl and 10Na:12Cl systems, respectively (see Tab.~\ref{tbl:all}). In addition, the absolute charge around every ion is on average ~0.8 |$e$|, rather than exactly 1 |$e$|, as expected for ions in bulk solution. 

Notably, a negatively charged electrode and a charged water layer is also observed in the presence of nominally neutral electrolyte, i.e. when a balanced population of ions (10Na:10Cl) or no ions (0Na:0Cl) are present in solution (see Tab.~\ref{tbl:all} and SI). This is in clear contrast with the assumption in standard computational approaches that the electrode remains neutral in the presence of a neutral electrolyte, and should be considered when developing future explicit models for neutral Pt-water interfaces. 

A discussion about the suitability of the PBE functional to describe the metallic moiety of the Pt-water interface and the charge transfer to and from the electrolyte~\cite{Tkatchenko2014, Mori2008} can be found in SI.

 In the following paragraphs we will show that our methodology and model system also make possible to go beyond current models for the atomic structure of the Pt-water DL and reveal that
mass density, structure and orientation of the water layers in contact with the electrode strongly depend on the applied potential.

\subsubsection{Interfacial water structure and orientation} 

Current models describing the metal/water interface~\cite{Hodgson2009,Carrasco2009,Carrasco2012} atomic structure have 
every second water molecule oriented parallel to the surface and the other pointing one H atom either toward or away from the surface. In these models it is assumed that the electrode polarization is not associated to any mass redistribution in the water layer in contact with the metal electrode\cite{Norskov2004, Rossmeisl2007,Gross2014}.
Here we reveal that exactly the opposite is the case. Our most remarkable finding is the strong dependence on the applied potential of the mass density of first and second water layers in contact with the electrode.

In our calculations, first and second water layers in contact with the electrode are unequivocally individuated by two well-defined peaks in the electrolyte mass density profile along the direction perpendicular to the electrode surface (see Fig.~7 in SI). Fig.\ref{fig:combined} (c) shows that the intensities of the first O and H density peaks increase as the metal electrode becomes more positively charged, namely as we move from system 12Na:10Cl, to 10Na:10Cl, to 10Na:12Cl. The analysis of these density peaks also show that all the water molecules in the first layer have, on average, almost the same orientation in every system studied, and lie on a plane nearly parallel to the metal, with the O atom nearer to the metal, and the H atoms slightly out of plane, pointing towards the bulk water. Thus, as the negative charge on the electrode decreases, a higher number of similarly oriented and positively charged water molecules is present in the first water layer in contact with it (see above and Tab.~\ref{tab:nb_water}).

 The mass density of the second water layer in our systems (see Fig.~\ref{fig:combined} (d)) is less affected by the electrode charging state. However, here water orientation strongly depends on the electrode charging state. The mass density profile for this water layer is characterized by broader density peaks and shows that one H atom per molecule is always located almost in the same plane, defined by the O atoms, and the other always points towards the Pt surface. As the negative charge on the electrode increases, water molecules progressively reorient, pointing a higher number of H atoms towards the metal electrode. Note that our electrodes are all negative.

To complement this picture, our analysis shows that the O atoms in the first layer form intralayer chemical bonds with the H atoms and interlayer H bonding, whose density depends on the electrode state, as highlighted by a constant coordination number of only $n$=2 under $1.5$ \AA, and $n$=2.2-2.8 at larger distances, up to 3~\AA (See Fig.\ref{fig:combined}(b)).  
%

\begin{figure}
 \includegraphics[width=0.8\textwidth]{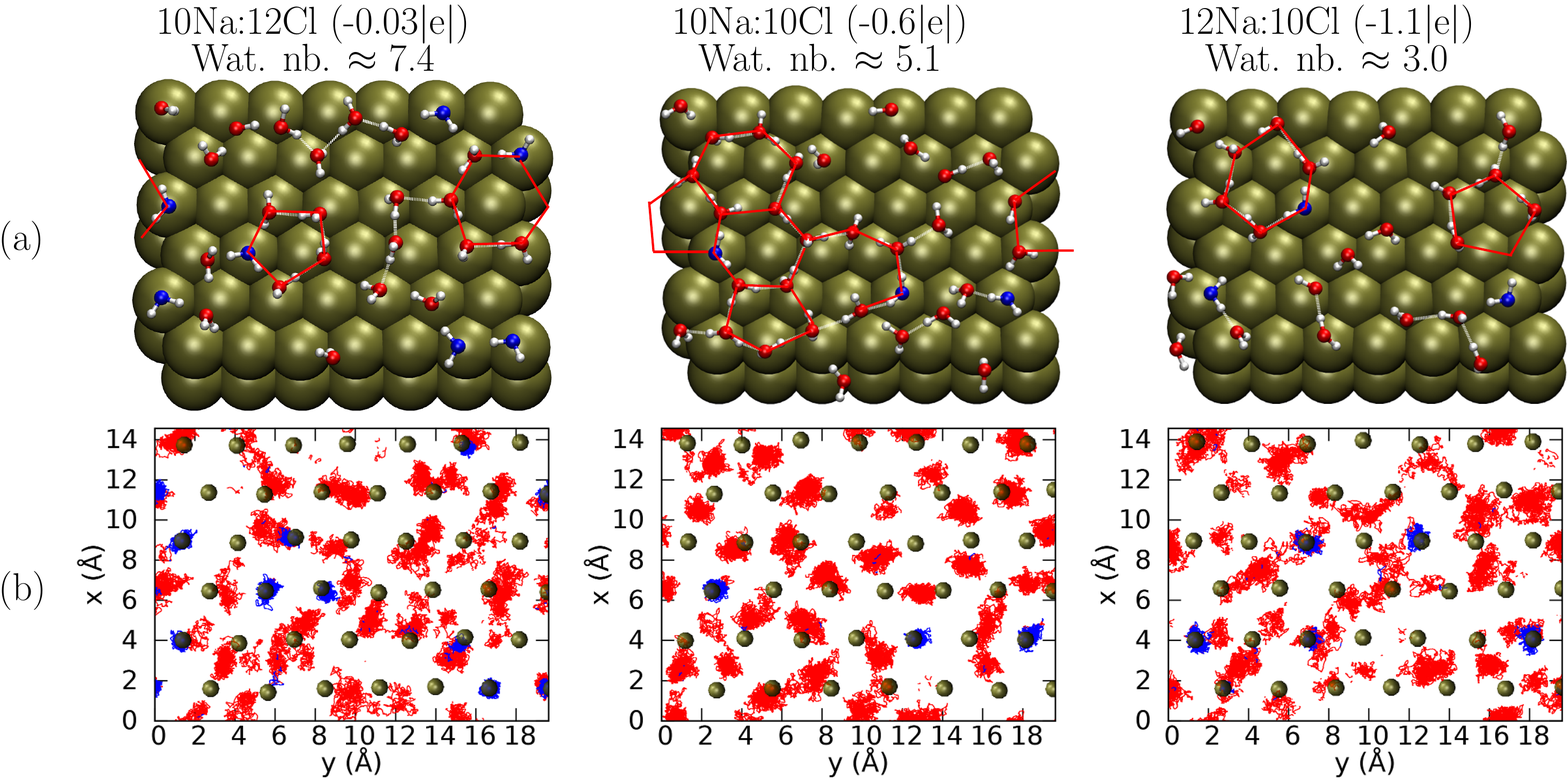}
 \caption{(a) Prototypical configuration of first and second water layer for 10Na:12Cl, 10Na:10Cl and 
 12Na:10Cl systems, from left to right. Red lines highlight hexagonal and pentagonal water motives on the Pt surface. (b) Trajectories of O atoms belonging to first and second water layers 
 in contact with the electrode, marked with blue and red lines, respectively. The charge on the metal 
 electrode is also reported (units of |$e$|). Wat. nb. indicates the average number of molecules in the 
 first water layer.}
  \label{fig:top_layers}
\end{figure}

Finally, we studied the time evolution of the position of the water molecules in contact with the electrode at different potentials. The positions of the O atoms belonging to first and second water layers are reported as a function 
of time in Fig.~\ref{fig:top_layers}(a). This figure clearly shows that the water molecules belonging to the first 
layer have their O atoms on the atop sites of the Pt(111) surface (blue traces in Fig.~\ref{fig:top_layers}) and do not diffuse within the time scale of our simulations ({$\approx$} 50 ps). In contrast, the water molecules belonging to the second water layer are more mobile, as shown by the red traces in Fig.~\ref{fig:top_layers}.

We also observed that interfacial water is arranged in hexagonal and pentagonal water rings, which are stable on Pt surface over the simulation time and have a different densities and distributions, depending on the electrode charging. Such water rings are formed by molecules belonging to the first and second water layer and are present even if the symmetry of the simulation supercell we used is not compatible in all directions with the formation of a closely packed hexagonal (or pentagonal) water arrangement on the surface. The observation of hexagonal and pentagonal water rings on Pt is in agreement with other works~\cite{Schnur2009, Meng2002,Nie2010,Sakong2015} and has been detected experimentally~\cite{Nie2010, Carrasco2009,Glebov1997}. We show here that the structure of the water layers near the Pt electrode (see Fig.~\ref{fig:top_layers}(a)) is potential dependent and results from the balance between the mass density of the first charged water layer and the orientation of the second one.

%
%

The results presented above establish a new model for the nanostructure of the DL at metal/water interfaces, which goes beyond both the standard Gouy-Chapman-Stern capacitor picture and current computational models. 
We have revealed a complex dependence of interfacial charge screening and polarization as well as atomistic structure and dynamics on the applied potential. We expect that our model will represent the baseline for future more realistic evaluations of kinetic 
overpotentials associated to electrocatalytic transformation at the Pt-water interface of huge technological relevance.

Our model for the Pt-water DL is new and is different from existing models; it is nevertheless consistent with the experimental observation of key EC quantities such as the interface capacitance, the electrode potential and the potential of zero charge. In the following, before illustrating how we computationally evaluated such important EC quantities, we will show how our methodology can be used to predict SFG spectra, which cannot be easily extracted experimentally.

\subsubsection{Computational Sum Frequency Generation Spectra}
The water orientation at the interface could in principle be insightfully probed by interface sensitive spectroscopy, such as SFG. Despite a couple of pioneering attempts to investigate the Pt-water interface with SFG~\cite{Backus2012, khatib_CaF2}, the experimental investigation is made challenging by the the presence of a strong non-resonant signal from the metal, which makes it impossible to directly access the water structure and orientation~\cite{Backus2012}. However, calculating the resonant SFG spectra at the Pt-water interface is computationally possible, and is here obtained evaluating the surface sensitive Vibrational Density of States (VDOS)~\cite{Khatib2017,khatib_CaF2} (see Methods section for details).
Using this approach, we observe in the first adsorbed layer (Fig.~\ref{fig:combined}(e)) a single positive band located at around 2800--3000~cm$^{-1}$, whose intensity increases as the positive charge on the electrode increases. The more intense the signal, the higher the density of the water molecules pointing towards the water bulk. The surface sensitive VDOS for the second adsorbed water layer, shown in  Fig.~\ref{fig:combined}(f), has a completely different behaviour; for all the interfaces studied, we find that a positive and a negative band are present. We observe that as the negative charge on the electrode decreases, the intensity of the positive band at lower frequencies increases and the negative bend is reduced. These trends indicate a higher density of dipoles reoriented as in the first water layer.

These VDOS data are consistent with the results presented above on mass density distribution and molecular reorientation under bias 
(illustrated in Fig.~\ref{fig:combined}(c-d) ) and provide a useful tool to experimentalists to 
distinguish and interpret the geometric structures of water at Pt interfaces under bias. 
The positive band at 2800-3000 is also in agreement with some very recent calculations by Cheng of simple VDOS, 
which showed similar peaks for the water molecules directly adsorbed on the Pt(111) surface\cite{Cheng2018}.
Cheng's calculations were performed at the PZC and the significant redshift observed, as compared to bulk water, was attributed to the combined effects of charge transfer and strong H-bonds.

\subsection{The Interface Capacitance}

We will now illustrate how with our \textit{ion unbalance} methodology we directly evaluated the capacitance, the zero point charge and the absolute electrode potential for each interface.

In our methodology the capacitance $C=  \Delta Q / \Delta V $ of the Pt-water interface is obtained \textit{a posteriori}, after all the equilibrium AIMD trajectories are produced, from the evaluation of the average interfacial potential drops, $\Delta V$, associated to the average DL charges, $\Delta Q$, for each charging state of the electrode.
 The DL charges are here evaluated from Bader analysis and include for each system the charge on the electrode and in the charged water layers in its proximity; the associated potential drops can be compared 
 as in our systems the bulk electrolyte condition is reached in the middle of each electrolyte region, and therefore the associated potential energy level can be used as a common reference; this bulk electrolyte condition is clearly marked in all our systems by the locally flat average electrostatic potential energy and density profiles; and by a locally neutral electrolyte solution in the centre of each cell (see Fig.~1 in SI).

\begin{figure}
  \includegraphics{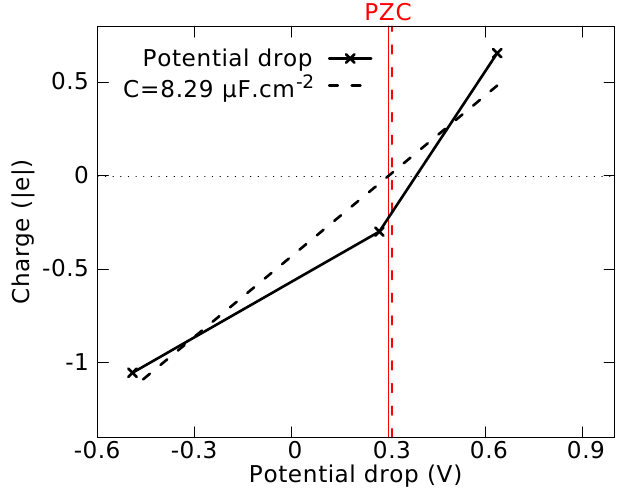}
  \caption{Double layer charge versus potential drop $\Delta V$ , as averaged along the trajectory for 12Na:10Cl 
  (electrode charge -1.1 |$e$|), 10Na:10Cl (electrode charge -0.6 |$e$|) and 10Na:12Cl (electrode charge -0.03 
  |$e$|) systems. Charge and potential drop are provided in units of |$e$| and Volts, respectively. Also reported, 
  the system's capacitance ($C= 8.29~\mu$F$\cdot$cm$^{-2}$), evaluated from the slope of the charge/potential 
  linear fit (dashed black line). The continuous vertical red line represents the experimental $PZC$. Aligning this 
  potential to the point where the linear fit crosses zero charge, allowed to define ad absolute potential scale for our 
  electrodes, referred to the standard hydrogen electrode potential. The dashed vertical red line represents our 
  evaluation of the $PZC$, where $\Delta V$ is evaluated relating the interfacial Fermi energy in every system to 
  the vacuum level, and this latter to the SHE. See SI for more detail on the evaluation of $\Delta V$. }
  \label{fig:Capacitance}
\end{figure}

In our calculations, the interfacial potential drops associated to different charging states of the electrode are clearly distinguishable during the simulation time and depend linearly on the interfacial charge (Fig.~3 and Fig.~4 of the SI). The capacitance is here calculated as the slope of the linear fit of the potential drop/charge relation 
(see Fig.~\ref{fig:Capacitance}) and amounts to $C= 8.29~\mu$F$\cdot$cm$^{-2}$. This is close to the 
experimental value~\cite{Mirtaheri2005} for the capacitance of a Pt-water electrolyte solution interface at 
high frequency and low NaCl concentration, which is expected to be lower than 10~$\mu$F$\cdot$cm$^{-2}$. Note that this experimental value does not include the low frequency modes determined by the presence of the diffused layer~\cite{Pajkossy2001}.

 From the calculated potential drop trend we also straightforwardly determined the {\it{potential of zero charge (PZC)}} for our Pt-water interface, by linearly fitting our data and finding the potential corresponding to the zero charge point. By aligning this potential value to the experimental~\cite{ZeroChargeJiabo, Cuesta2004} $PZC$ for a Pt-water interface, 

 it is then possible to define an absolute scale for the electrode potential in our systems. In this way, the absolute electrode potentials with respect to the SHE of the 12Na:10Cl, 10Na:10Cl and 10Na:12Cl systems are calculated at -0.48, 0.27 and 0.64 V, respectively. 
  Note that to evaluate the capacitance with our methodology, there is no need to place the electrode potential on an absolute scale, as only differences between the potential drops corresponding to different electrode charges are needed.

Finally, we evaluated the $PZC$ computationally, by relating the potential drop to the vacuum potential, 
and using the linear fit procedure described above (see Fig.~3 in the SI). 
To evaluate the potential drop with respect to the vacuum potential we inserted a vacuum slab in the middle of 
cell and measured the difference between the Fermi level and the vacuum potential~\cite{Filhol2006}.
The $PZC$ with respect to the SHE calculated in this way amounts to $PZC= 0.31$ V, 
a value extremely close to the experimental one~\cite{ZeroChargeJiabo,Cuesta2004}. 
A discussion about possible different ways to evaluate the potential drop $\Delta V$, that are in principle equivalent, can be found in SI.

In Summary, we have presented the first realistic model for the nanoscopic structure of the Pt-water DL under applied potential. This was obtained by developing an \abinit methodology, the \textit{ion unbalance} methodology, which advances previous models towards a more realistic and dynamic description of EI, and uses simulations to carry out virtual experiments. Our methodology has enabled the direct observation of the nanoscopic structure of the DL under applied potential and revealed that electrode potarization is associated with a mass redistribution and a charge oscillation within the DL, both of which strongly depend on the applied potential.
We have shown that the metal/surface charging cannot be simply described in terms of the Gouy-Chapman-Stern capacitor model and have proposed a more complex picture, where the electronic charge distribution oscillates at the interface and spills over the aqueous electrolyte in contact with the metal. Furthermore we have demonstrated that 
the interfacial dipole under bias is not merely determined by reorientation of the first water layer in contact with the metal surface, but by its charging state, in combination with its number density, while water reorientation becomes relevant only in the second layer. Finally, our methodology has provided an intuitive approach to evaluate the electrode potential wrt the SHE and key EC quantities such as the potential of zero charge and the capacitance of the EI.

bsolute electrode potential and key EC quantities such as the potential of zero charge and the capacitance of the EI.  

The development of our comprehensive microscopic model for the DL under applied potential is extremely significant for fundamental chemistry, physics and interface science. Perhaps most importantly, it makes possible for the first time to design highly realistic models for catalytic processes under bias at interfaces of huge technological relevance. These processes play a key role e.g. in solar, chemical fuel production and fuel cell technology. Thus, our results pave the way for developing new energy transformation models.

\section{Methods}
\subsection{DFT-MD}
MD based on DFT is performed for all systems using the Quickstep module of the CP2K package~\cite{quickstep}. 
The electronic structure is obtained at the PBE level~\cite{PBE}. Tether, Goedecker and Hutter (GTH)
pseudopotentials~\cite{gth1}, valence triple-zeta TZV basis set for Pt, TZVP (TZV polarization)
basis sets for the other atoms and a cutoff energy of 300 Ry for the plane waves are used. 
For every atomic configuration and every system studied, the potential energy is computed by minimising the 
electronic DFT functional, while time evolution is simulated by Born-Oppenheimer MD, using 
the gradients of the DFT potential energy surface to provide the forces entering Newton's equations of motion. 
The time-step used to integrate the equations of motion is 0.7~fs within the NVT ensemble. The temperature is 
controlled by using a Langevin thermostat~\cite{kress2003,galli2004,parrinello2004} set at 340~K. 
For all systems PBC are used. The in-plane cell parameters are $a=14.568$~\AA and $b=19.626$~\AA. 
In order to keep the density of the aqueous solution consistent with the studied system ($\approx$ 1 kg$\cdot$L$^{-1}$ 
or slightly higher, depending on the salt concentration), the normal axis, $c$ changes in the interval $35.942$ \AA 
$\leq c \leq 39.603$ \AA.

The Pt-water half-cell is realistically modelled with nearly 1000 atoms. In particular, we have 768 Pt atoms, 250 
water molecules, a variable number of  \ce{Na+} and  \ce{Cl-} ions, depending on composition, and about 5250 electrons. 
The Pt(111) electrode surface is modelled by using a 4-layer slab (42 Pt atoms per layer).  The coordinates of the 
central layers of the Pt slab are constrained to the bulk electrode values. Using a four layer slab introduces an error 
in the evaluation of the Pt work function of less than 0.04~eV. This is acceptable in view of the reduced computational 
cost. Extensive tests have been performed over the slab size against sampling of the Brillouin zone.

\subsection{Assessment of the \textit{Ion unbalance} methodology}

The analysis of the Kohn-Sham density of states projected on the metal moiety and the Na and Cl ionic species in our systems (PDOS, represented in Fig.3 in SI), confirms that the position of the ions centred energy levels with respect to the Pt Fermi level, $E_\mathrm{F}$, is qualitatively correct, as it reproduces the one schematically represented in Fig.~\ref{fig:Levels}. In particular, the lowest unoccupied energy level on \ce{Cl} is below the Pt Fermi level, while the highest occupied energy level centered on \ce{Na} is above it. This indicates that electrons localise around the \ce{Cl} atom, forming a \ce{Cl-} anion and leave \ce{Na+} in the cationic form. Note that we did not observe any water centred states in the gap between the ion centred energy levels; this confirms that the correct energy alignment is qualitatively reproduced by DFT in our systems.

We also observe that the PDos on Cl and Na ions in the three systems studied (12Na:10Cl, 10Na:10Cl and 10Na:12Cl) are aligned. Thus, even if the three interfaces have slightly different ionic composition, their ions in solution have a common status of charge and can be considered as three expressions of the same Pt-water interface under different bias conditions.

Finally, we find that the structure and the charge of the ion solvation shells in our systems only depend on the nature of the solvated ions, and are not affected by varying the  number of ions in solution. This further confirms that creating an unbalance of ions in the solution (adding or subtracting few ions) can be used as a strategy to 
model different electrode states which does not affect the electrolyte properties. Note that in every system studied 
the overall charge associated to the first solvation shells around Na and Cl ions is nearly equal in magnitude, but opposite 
in sign (see below for a full picture of Bader charges in our system). Thus, when an equal population of anions and cations 
is present in solution their overall charge almost cancels out (see Tab.\ref{tbl:all}).

\section{Data Availability}
Authors can confirm that all relevant data are included in the paper and/ or its supplementary information files.


\begin{acknowledgement}
This work was enabled by EPSRC (EP/P033555/1), Science Foundation Ireland (SFI) funded centre AMBER (SFI/12/RC/2278) and Irish Research Council (IRC) postgraduate grant GOIPG/2014/1392, DFG Research Grant SU 752/2-1,TRR146. The calculations performed for this work were performed using ARCHER, UK National Supercomputing Service (http://www.archer.ac.uk, via our membership of the UK's HEC Materials Chemistry Consortium, which is funded by EPSRC (EP/R029431)), the Kelvin and Boyle clusters, maintained by the Trinity Centre for High Performance Computing (project id: HPC-16-00932, these clusters were funded through grants from Science Foundation Ireland) and HRLS, Stuttgart, DE. We gratefully acknowledge Prof. Pietro Ballone for helpful discussions about the \textit{ion unbalance} methodology and Dr. Akinlolu Akande, for assistance with some figures at an early stage of the preparation of this paper.
\end{acknowledgement}

\section{Author Contributions}
CSC wrote the manuscript with RK contributing and all authors contributing to a minor extent. RK, AK and CSC analysed the data with contribution from MS. RK made the pictures with contribution from CSC, performed ss-VDOS analysis with contribution from MS, and calculated density profiles with contribution from AK. AK performed Bader analysis with contribution from RK and CSC. All authors discussed the data. CSC performed the AIMD simulations and guided and supervised the project.

\begin{suppinfo}

List of the contents of SI:

\begin{itemize}
\item Bulk electrolyte condition
\item Evaluation of the electrode/electrolyte potential drop
\item Ability of PBE functional to describe qualitative energy level alignment, the metallic moiety of the Pt-water interface and the density of states of the Ag-water solution interface
\item Water structure and orientation
\item ss-VDOS 
\item Bader charges
\item Van der Waals interactions
\end{itemize}

\end{suppinfo}

\bibliography{biblio}

\end{document}